\documentclass[aps,prd,reprint,groupedaddress,showpacs,nofootinbib,amsmath,onecolumn]{revtex4-1}

\usepackage{amsmath}
\usepackage{amssymb}
\usepackage{bbold}
\usepackage{graphicx}
\usepackage{verbatim}
\usepackage{hyperref}
\usepackage{indentfirst}
\usepackage{mathtools}
\usepackage{xcolor}
\usepackage{color}
\usepackage{adjustbox}
\usepackage{placeins}
\usepackage{lipsum}
\usepackage{csquotes}

\usepackage{float}
\restylefloat{table}

\usepackage{soul}

\usepackage{tikz}

\usepackage[font=small,labelfont=bf,labelsep=space]{caption}
\begin{document}

\title{Inflationary epoch in the presence of holographic dark energy}

\author{Paola C. M. Delgado}\email{paola.moreira.delgado@doctoral.uj.edu.pl}
\affiliation{Faculty of Physics, Astronomy and Applied Computer Science, Jagiellonian University,
30-348 Krakow, Poland.}

\author{Alexander Ganz}\email{alexander.ganz@uj.edu.pl}
\affiliation{Faculty of Physics, Astronomy and Applied Computer Science, Jagiellonian University,
30-348 Krakow, Poland.}

\author{Chunshan Lin}\email{chunshan.lin@uj.edu.pl}
\affiliation{Faculty of Physics, Astronomy and Applied Computer Science, Jagiellonian University,
30-348 Krakow, Poland.}

\date{\today}

%\date{\today}

\begin{abstract}
We analyze the effects of the holographic dark energy model in a single field slow-roll inflation, taking into account both the holographic and the dark radiation components. In particular, we obtain the background evolution and compute the scalar and tensor power spectra. For the scalar sector we show that the power spectrum of the curvature perturbation encompasses the standard single field result and a correction proportional to $\Omega_{\rm hde}/\epsilon$, where $\Omega_{\rm hde}$ is the fractional density of the holographic component and $\epsilon$ is the first slow-roll parameter. This correction might be of order unity in the very beginning of the inflationary phase and decays rapidly. For the primordial gravitational waves we find the spectral index receives a correction from the graviton mass term, which decays in the first inflationary e-folds.
\end{abstract}

\maketitle

%\tableofcontents

\section{Introduction}\label{Introduction}

In 1998, observations of type Ia supernovae led to the conclusion that the Universe is under an accelerated expansion \cite{Riess,Perlmutter}, which gave rise to the concept of dark energy \cite{PeeblesDE}. In the following years, the presence of this dark component was corroborated by many other observations, such as the Cosmic Microwave Background (CMB) radiation \cite{WMAP,Planck2013,Planck2015, Planck2018}, the large scale structure \cite{LSS}, the late-time integrated Sachs-Wolfe effect \cite{SWE1,SWE2} and the direct measurement of the Hubble parameter \cite{measurementsHubble}. However, althought its existence is widely evidenced, the nature of dark energy remains unknown. The standard cosmological model, namely $\Lambda{\rm CDM}$, is currently favoured by observations \cite{DES1, DES3}, meaning that a cosmological constant $\Lambda$ would be  responsible for driving the accelerated expansion. 

On the other hand, the holographic principle \cite{HP1,HP}, which states that the physics inside a volume can be described by a theory on its boundary, has been considered as a principle of quantum gravity and, therefore, could shed some light on the dark energy problem.  According to the holographic principle, the vacuum
energy density arising from the quantum fluctuation of the UV cut-off quantum field theory should relate to the boundary surface of a system in the way \cite{Cohen:1998zx}, 
\begin{equation}\label{hp and da}
	\rho\propto M_p^2 L^{-2},
\end{equation}
where $M_p$ is the reduced Planck mass and $L$ is a length scale. On the UV side, the quartic divergence of the vacuum energy is cut off at the scale $\Lambda \sim M_p/L$.  Therefore, it can serve as one of the solutions to the cosmological constant problem \cite{Weinberg:1988cp}.  It was suggested to adopt the future event horizon $R_h$  as the IR cut-off scale of the universe  \cite{MiaoLi1}, and the cosmic expansion is speeded up by the associated  vacuum energy given by eq.~(\ref{hp and da}).  The model is dubbed as holographic dark energy (HDE), and has drawn a lot of attention, being widely
studied. See Ref.  \cite{MiaoLi2} for a comprehensive review on the topic.

Recently, a general covariant local field theory of HDE was presented in \cite{ChunshanHDE}, where it is shown that the low energy effective theory corresponds to a massive gravity, whose graviton has 3 polarisations, including 1 scalar and 2 tensor. The UV cut-off of the HDE stems from the strong coupling nature of the scalar mode above some certain energy scale that relates to the graviton's Compton wavelength. It provides a physical interpretation for the UV-IR correspondence of HDE. 

Given the general covariant local field theory of HDE, it is thus possible to analyse its local dynamics. The structure formation in the framework of the effective field theory of HDE was studied in the Ref. \cite{Ganz:2021hmp}, and it has been found that the equation of motion for the matter density contrast $\delta_m\equiv \delta\rho_m/\rho_m$ of the cold dark matter is the same as the one in the general relativity up to the leading order in the small scale limit $k\gg aH$, provided the equation of state is Quintessence-like. 

Since HDE is present during the whole cosmic history, including the early Universe, it is expected that it alters the dynamics during inflation \cite{Inflation1,Inflation2,Inflation3,Inflation4,Inflation5}, which is the most accepted solution to the horizon and flatness problems in cosmology, encompassing models in great agreement with the observed scalar power spectrum and with the not yet detected primordial gravitational waves. It has been pointed out that the the cosmic coincidence problem can also be solved, provided a minimal number of e-folds during inflation \cite{MiaoLi1}.  The corrections of HDE to the primordial curvature perturbations were analysed in \cite{InflationHDE}, leading to the conclusion that the scalar power spectrum is generically red-tilted. However, this analysis is incomplete as the local field theory of HDE was still missing at that time, and the analyses did not include the perturbations of HDE itself. In the present work, with the general covariant and ghost free action obtained in \cite{ChunshanHDE} in hands, which takes into account also the contribution of perturbations of HDE, we compute the inflationary background and both scalar and tensor power spectra\footnote{In \cite{HDEinflationdifferent} the authors propose that HDE is the responsible for driving the inflationary phase, which differs from the approach developed in the present work.}. 

The paper is organized as follows: in section \ref{Validity} we discuss the validity of the low energy effective field theory obtained in \cite{ChunshanHDE} during the inflationary phase. In section \ref{Inflationary background} we obtain the background evolution of the inflaton and of the dark components numerically. In section \ref{Cosmological perturbations} we compute the cosmological perturbations on a Friedmann-Lemaître-Robertson-Walker background and present the quadratic actions for the scalar and tensor parts. In section \ref{Power spectra} we compute the scalar and tensor power spectra and discuss their compatibility with observational constraints. Finally, in section \ref{Conclusions} we summarize the results and discussions. 

\section{Validity of the effective field theory during the inflationary phase}\label{Validity}
In obtaining equation \eqref{hp and da} via the holographic principle, one sets an UV cut-off $\Lambda$ to the local quantum field theory \cite{MiaoLi2} and a UV-IR correspondence takes place. In essence, one requires the energy within a Schwarzschild radius, i.e. $L^3\Lambda^4$, to be smaller than the mass of a corresponding black hole, i.e. $L M_p^2$, so that $\Lambda\lesssim\sqrt{M_p/L}$. As mentioned in section \ref{Introduction}, the IR cut-off is chosen to be the future event horizon $R_h=aL$, so that $\Lambda\sim\sqrt{M_p/R_h}$. Since the later varies in time, the UV cut-off $\Lambda$ is also time dependent. 

The relevance of the holographic dark energy to the cosmological scenario depends on whether $\Lambda>H$ is satisfied, where $H$ is the Hubble parameter. Recalling that $\Omega_{\rm hde}\propto 1/(H^2 R_h^2)$, where $\Omega_{\rm hde}$ is the fractional density of the holographic term, one finds that this condition can be rewritten as
\begin{equation}\label{validity condition}
    \Omega_{\rm hde}>\frac{H^2}{M_p^2}.
\end{equation}
As we are going to show in section \ref{Inflationary background}, during inflation the holographic component constitutes a tiny amount of the total energy density, i.e. $\Omega_{\rm hde}\ll 1$. Recalling that it scales as $a^{-2}$, we see that for large field inflation, for which $H^2/M_p^2\sim 10^{-12}$, the effective field theory breaks down  some time around $10$ to $12$ e-folds after the beginning of inflation, which is the moment when the inflaton energy density dominates over the one of HDE.  Although this time interval is quite limited, it is enough to cover the CMB scales. For small field inflation, the ratio $H^2/M_p^2$ is much smaller and, therefore, the effective theory has a much longer validity. 

In what follows we assume that the inflationary phase continues even after the break down of the effective theory, since it does not necessarily mean a pathology, but rather that an UV completion is required.

\section{Inflationary background}\label{Inflationary background}
We start with the covariant and ghost free action of the holographic dark energy model, as obtained in \cite{ChunshanHDE}, and the action of the matter sector corresponding to the inflaton $\chi$:
\begin{eqnarray}\label{action}
   S=\int d^4 x\sqrt{-g}\left\{\frac{\mathcal{R}}{2}- \varphi^{-2}\left[(c+\lambda)Z +\lambda \partial^{\mu}\varphi \partial_{\mu}\varphi+\frac{3d}{8Z}\bar{\delta}Z^{ab}\bar{\delta}Z^{cd}\delta_{ac}\delta_{bd}\right]-\frac{1}{2}\partial_\mu \chi \partial^\mu \chi-V(\chi)\right\}.
\end{eqnarray}
Here $g$ is the determinant of the metric,
$M_p^2\equiv 8\pi G=1$, $\mathcal{R}$ is the 4 dimensional Ricci scalar, $V(\chi)$ is the inflaton potential and $c$ and $d$ are constants. The extra field $\varphi$ is a time-like Stueckelberg field, which arises to recover general covariance, together with other three space-like Stueckelberg fields $\phi^a$, where $a, b$ are indices in the field space described by the metric $\delta_{ab}$. 
$Z^{ab}$ is another building block of the theory, given by
\begin{equation}\label{Zab}
    Z^{ab}\equiv \partial_\mu \phi^a \partial^\mu \phi^b -\frac{(\partial_\mu \varphi \partial^\mu \phi^a)(\partial_\nu \varphi \partial^\nu \phi^b)}{\partial_\mu \varphi \partial^\mu \varphi}.
\end{equation}
As discussed in \cite{ChunshanHDE}, when the four Stueckelberg scalar fields assume their vacuum expectation values (VEVs), $\left \langle\varphi\right \rangle$ and $\left \langle \phi^a \right \rangle$, they break the diffeomorphism invariance. This introduces a massless boson for each broken symmetry, according to the Goldstone theorem. We then have $\varphi=\left \langle\varphi\right \rangle+\pi^0$ and $\phi^a=\left \langle \phi^a \right \rangle+\pi^a/\sqrt{3}$, where $\pi^0$ and $\pi^a$ are the time-like and space-like Goldstone bosons, respectively.
These Goldstone excitations satisfy the symmetry 
\begin{eqnarray}
\pi^i(t,\textbf{x})\to\pi^i(t,\textbf{x})+\xi^i(t),
\end{eqnarray}
which eliminates the dynamics of the 3 would-be ghosty bosons $\pi^i$. In the unitary gauge, the bosons are muted, and the graviton  becomes massive. With this gauge choice, eq.~\eqref{Zab} is rewritten as $Z^{ab}=h^{ij}\delta^a_i \delta^b_j/3$, where $h^{ij}$ is the induced metric of the spatial hypersurfaces in the ADM decomposition and $\delta^a_i$ is a pullback mapping between the space-time and the field space. The quantities $Z$ and $\bar{\delta}Z^{ab}$ are defined, respectively, as $Z\equiv Z^{ab}\delta_{ab}$ and $\bar{\delta}Z^{ab}\equiv Z^{ab}-3 Z^{ac}Z^{db}\delta_{cd}/(Z^{cd}\delta_{cd})$.  

In a Friedmann-Lemaître-Robertson-Walker background $ds^2=-dt^2+a^2d\textbf{x}^2$, the equations of motion read
\begin{eqnarray}
   3H^2&=&\frac{c}{a^2\varphi^2}+\frac{\lambda}{2a^4}+\frac{1}{2}\dot{\chi}^2+V(\chi),\\
   -\dot{H}&=&\frac{c }{3a^2\varphi^2}+\frac{\lambda}{3a^4}+\frac{1}{2}\dot{\chi}^2
\end{eqnarray}
and 
\begin{eqnarray}
  \label{phidot} \dot{\varphi}=-\frac{1}{a},\qquad \dot{\lambda}=-\frac{4 c a}{\varphi^3},
\end{eqnarray}
where overdot denotes derivative with respect to cosmic time $t$ and the Lagrangian multiplier was re-scaled as 
$\lambda\rightarrow \lambda \frac{\varphi^2}{4a^2}$ for simplicity. 
Note that the energy density of the dark sector includes two components,
\begin{equation}
	\rho_{dark}=\frac{c}{a^2L^2}+\frac{\lambda}{2a^4},
\end{equation}
where the first term is the holographic term, and the second term is the dark radiation term which  scales as radiation at late times, when $\lambda$ is approximately constant. For this reason, from now on we refer to the above contributions as the holographic term $\rho_{\rm hde}$ and the dark radiation term $\rho_{\rm rad}$ respectively.

In order to obtain the background evolution numerically, we define the following dimensionless quantities, in analogy to \cite{MiaoLiHDE}:
\begin{eqnarray}
  \tilde{\varphi}\equiv H_i\varphi,\qquad \tilde{\lambda}\equiv \frac{\lambda}{H_i^2},\qquad
   E(z)\equiv\frac{H}{H_i},
\end{eqnarray}
where $H_i$ is the Hubble parameter at the beginning of the inflationary epoch. In terms of the above quantities, the Friedmann equation reads
\begin{equation}\label{E2}
    E^2=\frac{3H^2}{3H_i^2}=\frac{\rho_\chi+c a^{-2}\varphi^{-2}+\lambda(2a^4)^{-1}}{3H_i^2}=\tilde{\rho}_\chi+\frac{1}{3}\left(\frac{c }{a^2\tilde{\varphi}^2}+\frac{\tilde{\lambda}}{2a^4}\right),
\end{equation}
where $\rho_\chi$ is the energy density of the inflaton, $\tilde{\rho}_\chi\equiv\rho_\chi/\rho_c(t_i)$, and 
$\rho_c(t_i)\equiv 3H_i^2$ is the critical density at the beginning of inflation.

In the redshift space, equations \eqref{phidot} are rewritten as 
\begin{eqnarray}
\label{difphi}   \frac{d\tilde{\varphi}}{dz}=\frac{1}{E(z)},\qquad  \frac{d\tilde{\lambda}}{dz}=\frac{4c}{(1+z)^2E(z)\tilde{\varphi}^3},
\end{eqnarray}
while the equation of motion for $\tilde{\rho}_\chi$ is given by
\begin{equation}\label{difOmegachi}
    \frac{d\tilde{\rho}_\chi}{dz}=6(1+z)^{-1}(\tilde{\rho}_\chi-\tilde{V}),
\end{equation}
where $\tilde{V}\equiv V/\rho_c(t_i)$. For the latter the equation of motion reads
\begin{eqnarray}\label{difV}
   \frac{d\tilde{V}}{dz}=\frac{2\tilde{m}\sqrt{\tilde{V}(\tilde{\rho}_\chi-\tilde{V})}}{(1+z)E(z)},
\end{eqnarray}
where $\tilde{m}\equiv m/H_i$ and we have assumed the simplest chaotic inflation with $V=\frac{1}{2}m^2\chi^2$. 

\begin{figure}
\begin{center}
\includegraphics[width=7cm]{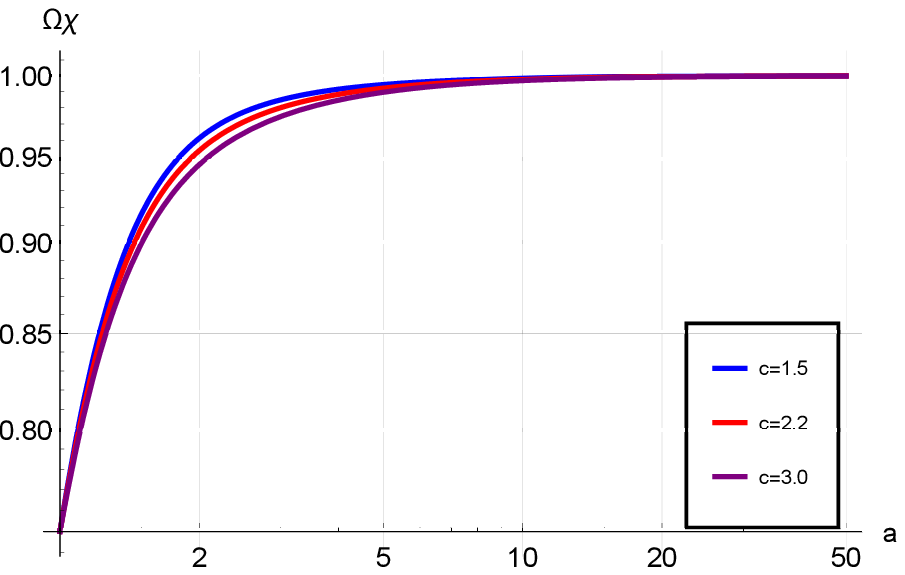}
\includegraphics[width=7cm]{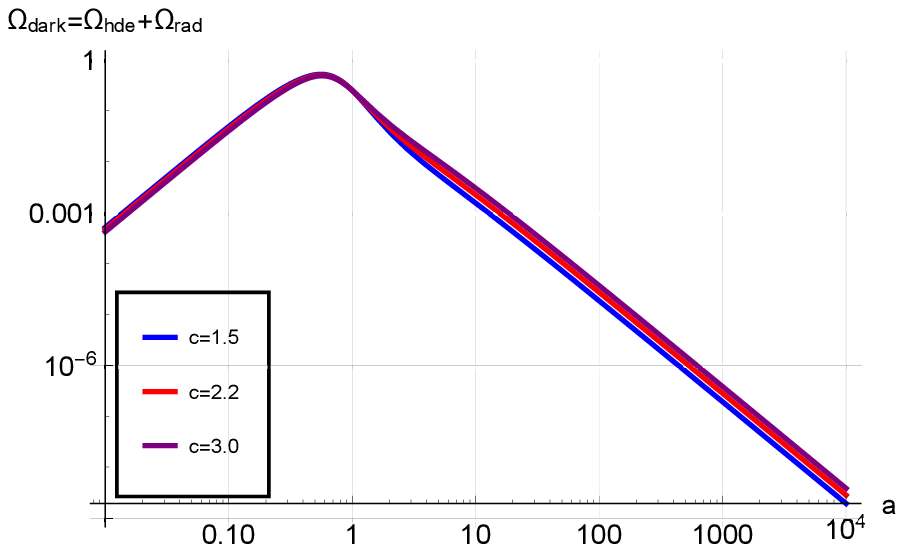}\\
\includegraphics[width=7cm]{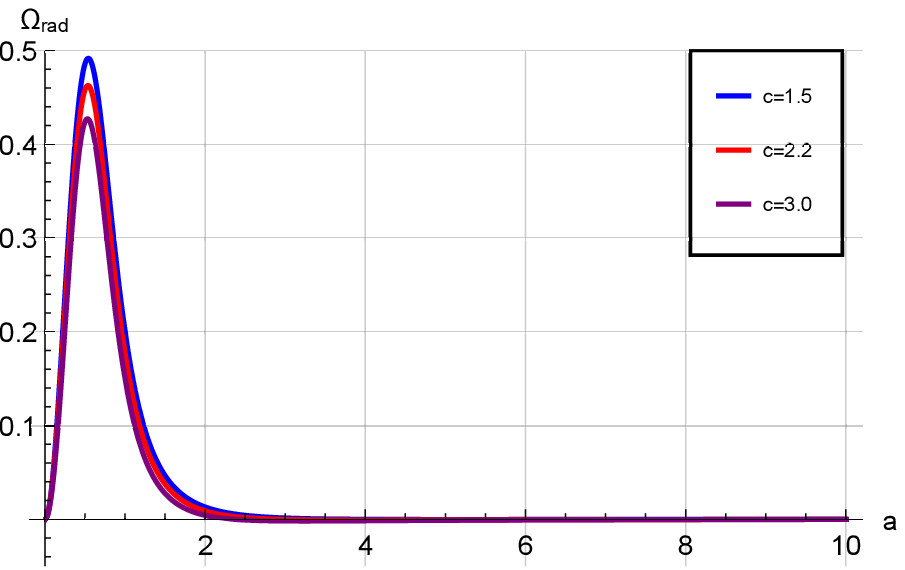}
\includegraphics[width=7cm]{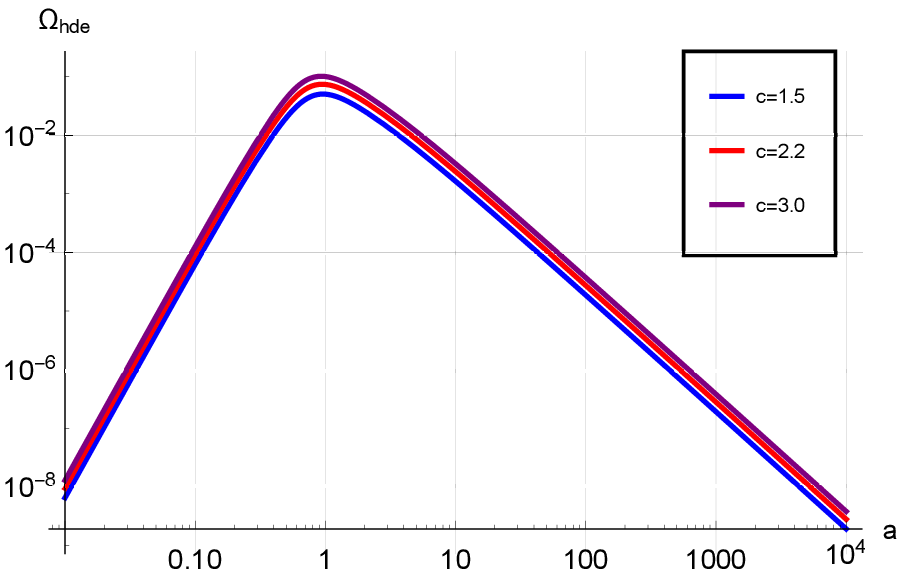}
\end{center}
\caption{\label{bg1} Evolution of the fractional energy densities of the inflaton $\Omega_\chi$, the dark radiation $\Omega_{\rm rad}$, the holographic component $\Omega_{\rm hde}$ and their sum  $\Omega_{\rm dark}$. The inflaton eventually dominates over the dark components after the beginning of inflation ($a_i=1$ in our convention). In these plots, $\tilde{m}=0.1$, $\tilde{\varphi}_i=\frac{1}{\sqrt{0.1}}$, $\tilde{\rho}_i=0.75$ and $\tilde{V}=0.7$.}
\end{figure}

\begin{figure}
\begin{center}
\includegraphics[width=7cm]{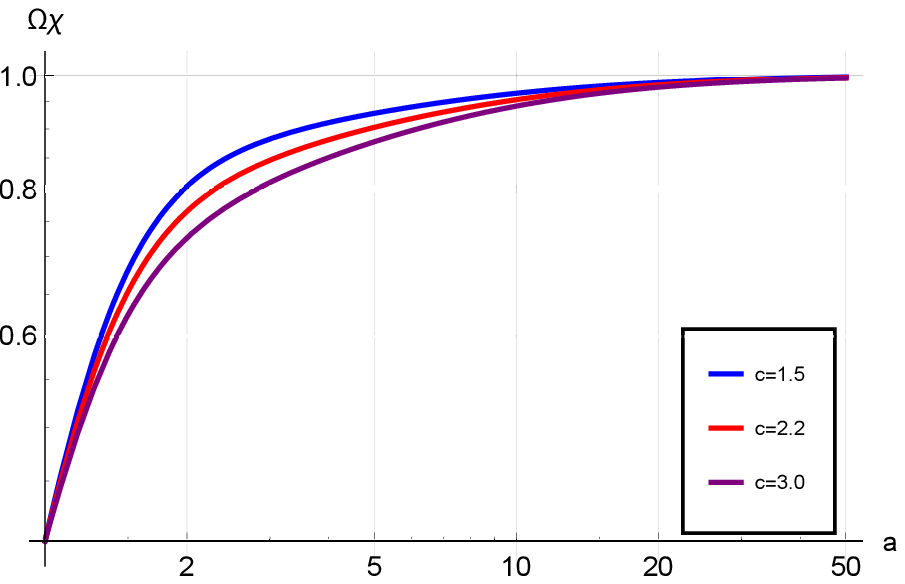}
\includegraphics[width=7cm]{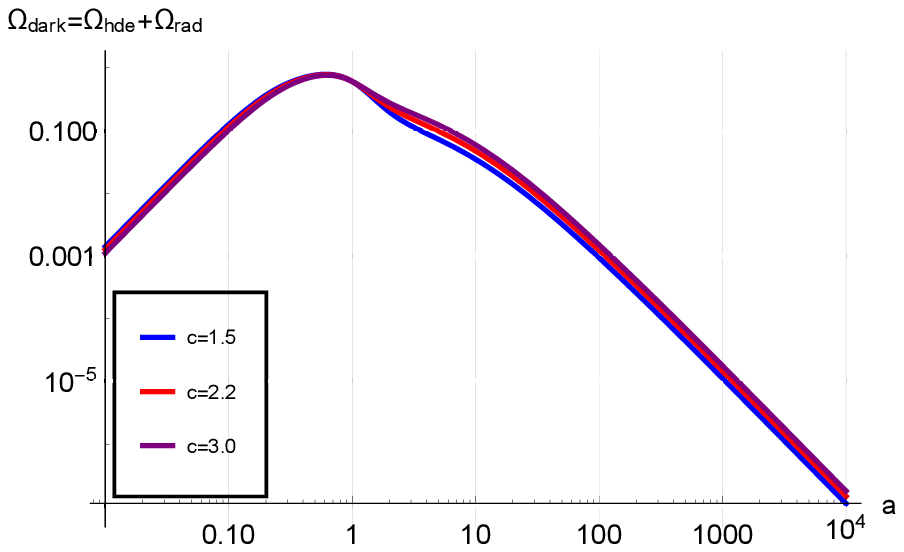}\\
\includegraphics[width=7cm]{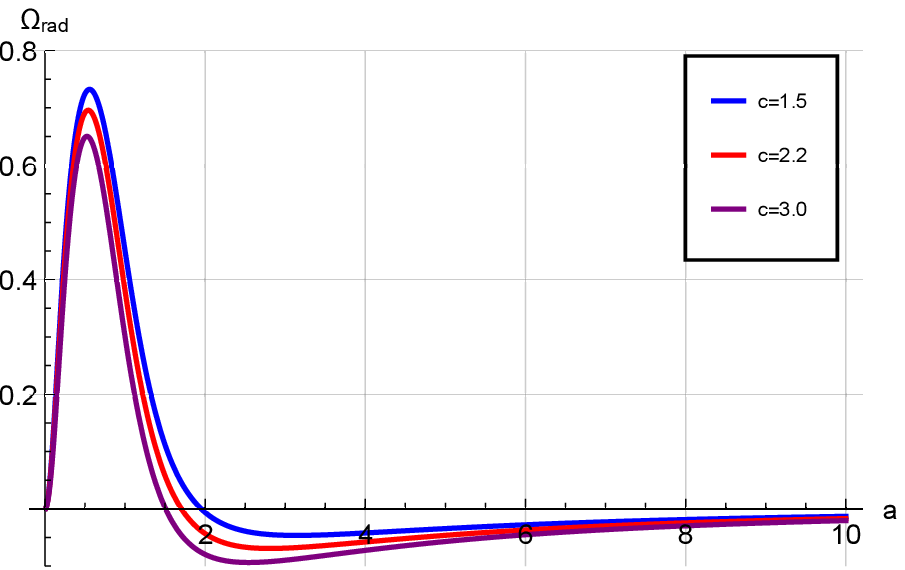}
\includegraphics[width=7cm]{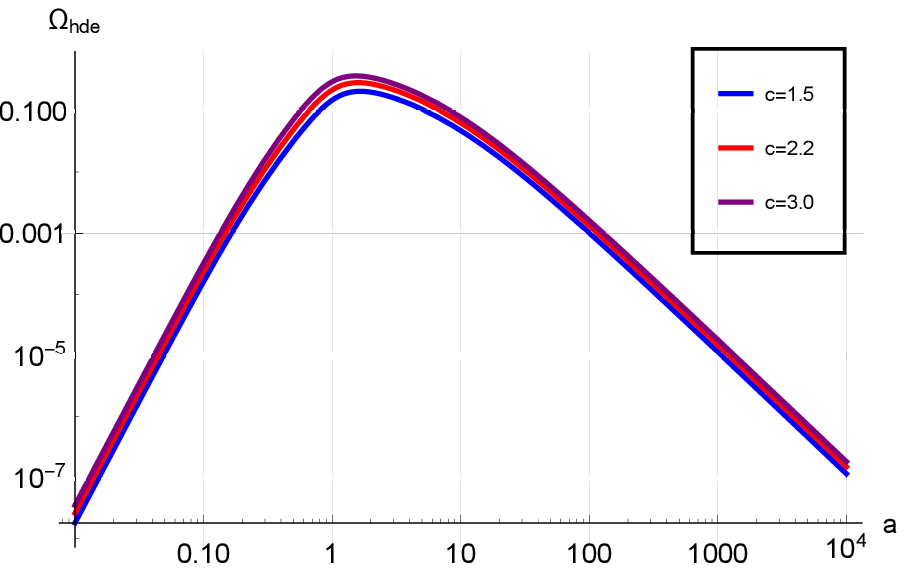}
\end{center}
\caption{\label{bg2} Evolution of the fractional energy densities of the inflaton $\Omega_\chi$, the dark radiation $\Omega_{\rm rad}$, the holographic component $\Omega_{\rm hde}$ and their sum  $\Omega_{\rm dark}$ with $\tilde{m}=0.1$, $\tilde{\varphi}_i=\frac{1}{\sqrt{0.3}}$, $\tilde{\rho}_i=0.4$ and $\tilde{V}=0.36$.}
\end{figure}

Numerically solving equations \eqref{E2}, \eqref{difphi}, \eqref{difOmegachi} and \eqref{difV} we obtain Fig.~\ref{bg1} and Fig.~\ref{bg2}, where 
we parametrized the scale factor as $a_i=1$ at the beginning of inflation. As initial conditions, we used the initial values of the time-like Stueckelberg field  $\tilde{\varphi}_i\equiv\tilde{\varphi}(0)$, the inflaton energy density $\tilde{\rho}_{\chi i}\equiv \tilde{\rho}_{\chi}(0)$, the inflaton potential $\tilde{V}_i\equiv\tilde{V}(0)$ and the Lagrange multiplier $\tilde{\lambda}_i\equiv \tilde{\lambda}(0)$ such that $E(z=0)=1$. The values of the constant $c$ were chosen according to the observational constraint obtained in \cite{MiaoLiHDE}, given by
\begin{equation}\label{bounds c}
    1.41<c<3.09
\end{equation}
with $95.4\%$ CL.
Note that the initial conditions determine the future event horizon $L$, and not the other way around. The constraint equation that allows us to obtain $L$ is given by $\dot{L}=-N/a$. We could, in principle, integrate it from $-\infty$ to nowadays
\begin{equation}\label{L1}
    L=\int_{-\infty}^{t}\frac{-N {dt}'}{a({t}')}+L(-\infty),
\end{equation}
where $L(-\infty)$ is the initial condition in the infinite past. However, this initial condition is not known and, in practice, we perform the integration from the other side: 
\begin{equation}\label{L2}
    L=\int_{t}^{\infty}\frac{N {dt}'}{a({t}')}+L(\infty).
\end{equation}
Equations \eqref{L1} and \eqref{L2} are equivalent, if both boundary conditions are known. In our case, $L(-\infty)$ remains unkonwn due to our ignorance about the quantum gravity. On the other hand, we do know the asymptotic value of $L(\infty)=0$, as it is model independent \cite{MiaoLi3}.

\begin{figure}
\begin{center}
\includegraphics[width=7cm]{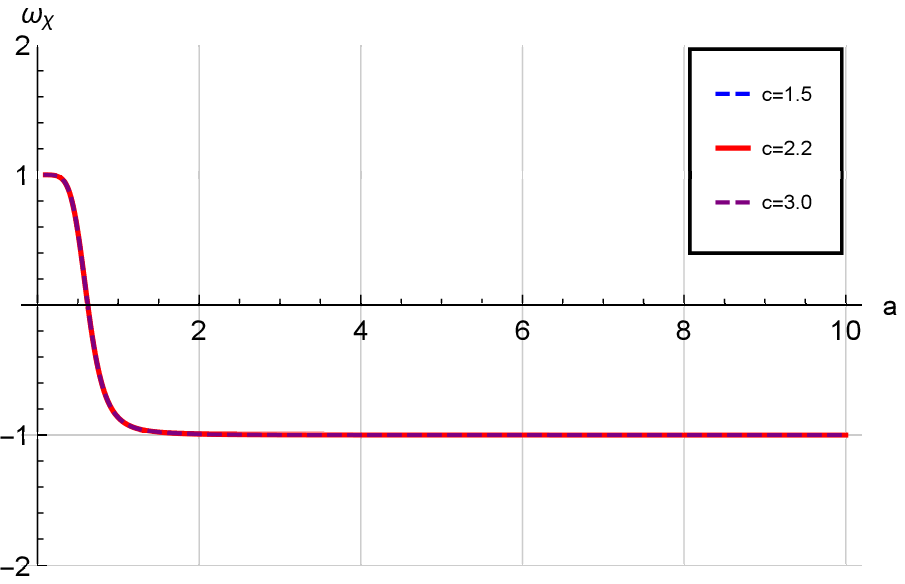}
\includegraphics[width=7cm]{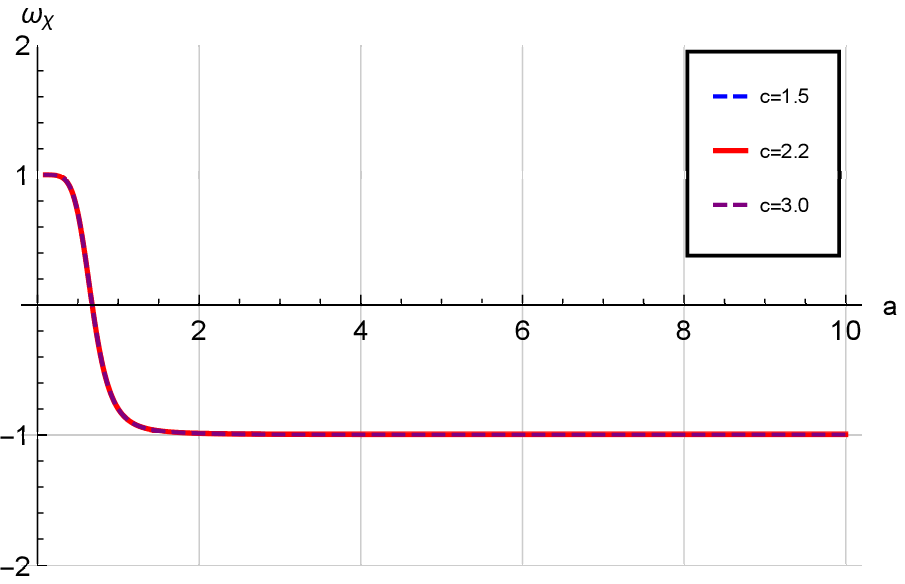}
\end{center}
\caption{\label{wchi} Equation of state parameter of the inflaton $\chi$ for $\tilde{m}=0.1$, $\tilde{\varphi}_i=\frac{1}{\sqrt{0.1}}$, $\tilde{\rho}_i=0.75$ and $\tilde{V}=0.7$ (left) and $\tilde{m}=0.1$, $\tilde{\varphi}_i=\frac{1}{\sqrt{0.3}}$, $\tilde{\rho}_i=0.4$ and $\tilde{V}=0.36$ (right). As expected, $\omega_\chi=-1$ in the inflationary phase.}
\end{figure}

From the numerical results we conclude that the inflaton eventually dominates over the dark radiation and holographic components, regardless of the hierarchy between them. In other words, at the background level the inflationary phase is not spoiled by the presence of the dark sector, making the holographic dark energy compatible with inflation. We can also see, from Fig.~\ref{bg2}, that $\Omega_{\rm rad}$ may become negative during the inflationary phase, but the density of the total dark sector $\Omega_{\rm dark}=\Omega_{\rm hde}+\Omega_{\rm rad}$ is nonetheless positive. 
The equation of state parameter of the inflaton, as expected, achieves $\omega_{\chi}=-1$ in the inflationary epoch, as shown in Fig~\ref{wchi}.

\section{Cosmological perturbations}\label{Cosmological perturbations}
In this section, the general covariant and ghost free action \eqref{action} is used to compute the cosmological perturbations in the unitary gauge around a Friedmann-Lemaître-Robertson-Walker background. We use the following metric decomposition 
\begin{eqnarray}
  \nonumber g_{00}&=&-a(t)^2(1+2\alpha),\\
 \nonumber g_{0i}&=&a(t)^2(\partial_i \beta+S_i),\\
  g_{ij}&=&a(t)^2\left[ \delta_{ij}+2\psi\delta_{ij}+\partial_i \partial_j E+\frac{1}{2}(\partial_i F_j +\partial_j F_i)+\gamma_{ij}\right],
\end{eqnarray}
where the scalar perturbations are represented by the variables $\alpha$, $\beta$, $\psi$ and $E$, the tensor perturbation is represented by $\gamma_{ij}$ and the vector perturbations are represented by $S_i$ and $F_i$. The latter obey $\partial_i S_i=0$ and $\partial_i F_i=0$, while $\gamma_{ij}$ obeys $\gamma_{ii}=\partial_i \gamma_{ij}=0$, being traceless and transverse. 
As mentioned in section \ref{Inflationary background}, in the unitary gauge we keep all the perturbation variables, while the Goldstone bosons are muted and the graviton acquires a mass.

\subsection{Scalar perturbations}\label{scalar perturbations}
For the scalar perturbations, the quadratic action reads
\begin{equation}
    S_{\rm scalar}^{(2)}=\int d^4 x~\left(\mathcal{L}_{\rm EH}+\mathcal{L}_{\rm mass}+\mathcal{L}_{\chi}\right),
\end{equation}
where $\mathcal{L}_{\rm EH}$ is the Einstein-Hilbert lagrangian density, $\mathcal{L}_{\rm mass}$ is the lagrangian density related to the graviton mass and $\mathcal{L}_{\chi}$ is the lagrangian density for the inflaton $\chi$. They read, respectively, 
\begin{eqnarray}
 \nonumber \frac{\mathcal{L}_{\rm EH}}{ a^3}&=&-3\dot{\psi}^2+\frac{k^2}{a^2}\left[\psi^2-\frac{1}{2}a^2\dot{E}(H\psi-2\dot{\psi})-\frac{1}{2}a^2HE(\dot{\psi}+3H\psi) \right]\\
 &+&\alpha \left[ \frac{2k^2\beta H}{a}+\frac{k^2}{a^2}(2\psi-a^2H\dot{E})+6H\dot{\psi}-3H^2\alpha \right]-\frac{2k^2\beta\dot{\psi}}{a},\\
\nonumber \mathcal{L}_{\rm mass}&=&\varphi^{-2}\left[-\frac{3c+d}{36}k^4aE^2-\frac{1}{6}ck^2aE(2\alpha+\psi)+ca\psi(2\alpha+\psi)\right]\\
&+&\lambda\left[-\frac{k^4E^2}{48a}+\frac{k^2E(\psi-\alpha)}{12a}+\frac{(\alpha+\psi)^2}{4a}\right]+\delta\lambda\left(\frac{\psi-\alpha}{2a}-\frac{k^2E}{12a}\right)
\end{eqnarray}
and
\begin{eqnarray}
 \nonumber \frac{\mathcal{L}_{\chi}}{a^3}&=&\frac{1}{2} \dot{\delta \chi}^2-\frac{1}{2}\dot{\chi}\dot{\delta \chi}(k^2E+2\alpha-6\psi)+\frac{1}{4}\dot{\chi}^2(2\alpha^2+k^2E\psi)\\
 &+&\delta\chi\left[-\frac{k^2\beta\dot{\chi}}{a}-\frac{1}{2}V_{,\chi}(-k^2E+2\alpha+6\psi)\right]+\delta\chi^2\left(-\frac{k^2}{2a^2}-\frac{V_{,\chi\chi}}{2}\right),
\end{eqnarray}
where the Lagrangian multiplier $\lambda$ was again re-scaled to 
$\lambda \frac{\varphi^2}{4a^2}$, overdot denotes derivative with respect to cosmic time and the sub-index $_{,\chi}$ denotes derivative with respect to the inflaton $\chi$. 

\begin{figure}
\begin{center}
\includegraphics[width=7cm]{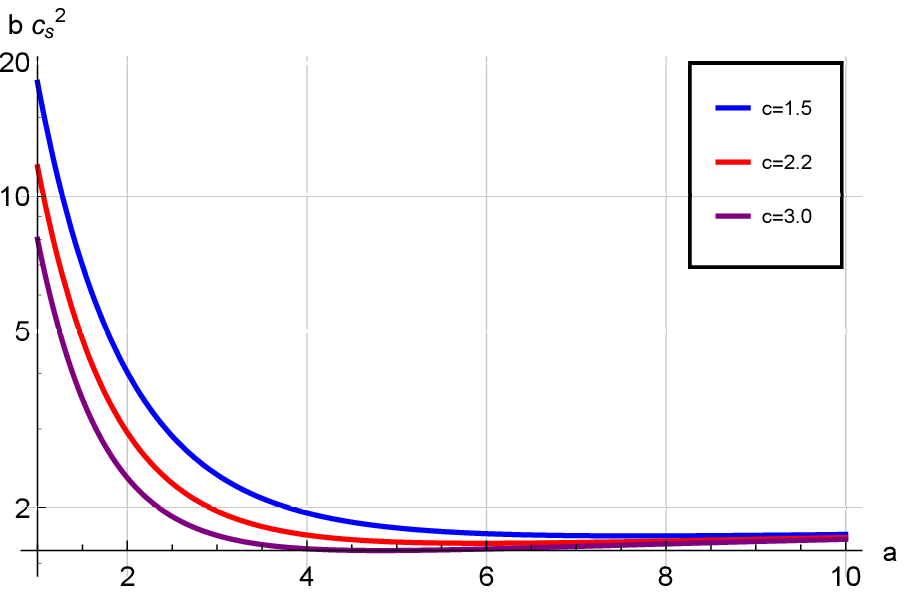}
\includegraphics[width=7cm]{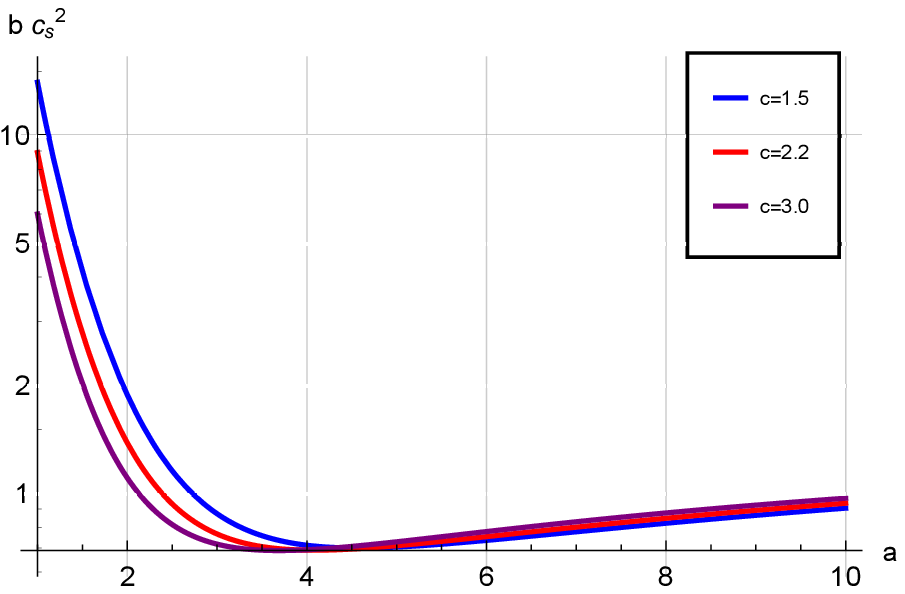}
\end{center}
\caption{\label{f:cs} Sound speed of the scalar graviton $\psi$ for $\tilde{m}=0.1$, $\tilde{\varphi}_i=\frac{1}{\sqrt{0.1}}$, $\tilde{\rho}_i=0.75$ and $\tilde{V}=0.7$ (left) and $\tilde{m}=0.1$, $\tilde{\varphi}_i=\frac{1}{\sqrt{0.3}}$, $\tilde{\rho}_i=0.4$ and $\tilde{V}=0.36$ (right).}
\end{figure}

The constraints can be obtained by varying the action with respect to the non-dynamical variables of the theory, namely $\alpha$, $\delta \lambda$ and $\beta$. We then have the following set of equations:
\begin{eqnarray}
\nonumber H(6\dot{\psi}-k^2\dot{E})+\frac{2c\psi}{a^2\varphi^2}+\frac{k^2}{a^2}\left(2\psi-\frac{cE}{3\varphi^2}\right)+\frac{\lambda}{12a^4}(6\psi-k^2E)&&\\
 -\frac{\delta\lambda}{2a^4}+\alpha\left(\frac{\lambda}{2a^4}-6H^2+\dot{\chi}^2\right)+\frac{2k^2H\beta}{a}-\dot{\chi}\dot{\delta \chi}-\delta\chi V_{,\chi}&=&0,\\
 k^2E+6\alpha-6\psi&=&0,\\
 -\frac{\dot{\chi}\delta\chi}{2}+H\alpha-\dot{\psi}&=&0.
\end{eqnarray}
Solving them simultaneously and substituting the solution back in the action, we find an expression of the form
\begin{equation}
    S_{\rm scalar}^{(2)}=\int d^4 x \left[\frac{a^3}{2}\dot{\delta\chi}^2+\frac{a^3\varphi^2\dot{\chi}^2-2(c+d)a }{2H^2\varphi^2}\dot{\psi}^2-\frac{a^3\dot{\chi}}{H}\dot{\psi}\dot{\delta\chi}+...\right],
\end{equation}
where the ellipsis stand for potential, gradient and (non-kinetic) interaction terms. The diagonalization of the kinetic terms is performed by redefining the $\delta\chi$ variable as follows
\begin{equation}\label{deltachioriginal}
    \delta\chi\rightarrow\delta\chi-\frac{C}{2A}\psi,
\end{equation}
where $A$ and $C$ are, respectively, the coefficients of $\delta \dot \chi^2$ 
and $\dot{\psi}\dot{\delta\chi}$. The quadratic diagonalized action then reads
\begin{equation}
    S_{\rm scalar}^{(2)}=\int d^4
x \left[-\frac{(c+d)a}{H^2\varphi^2} \dot{\psi}^2+\frac{a^3}{2}\dot{\delta\chi}^2+...\right],
\end{equation}
which requires $c+d<0$ to avoid ghost instabilities. Defining the constant $b\equiv-2(c+d)$ for convenience and the canonical variables $\delta\chi_c$ and $\psi_c$ as
\begin{eqnarray}
    \label{psic}\delta\chi_c\equiv a \delta\chi,\qquad\psi_c\equiv\frac{ \sqrt{b}}{H \varphi}\psi,
\end{eqnarray}
we rewrite the action as 
\begin{eqnarray}
    S_{\rm scalar}^{(2)}&=&\int dt d^3 k \frac{1}{2}\left[ a \dot{\psi_c}^2+a \dot{\delta\chi_c}^2+\left(-\frac{k^2 c_s^2}{a^2}-a M_s^2\right)\psi_c^2+\left(-\frac{k^2}{a^2}-a M_{\chi}^2\right)\delta\chi_c^2+I_1\delta\chi_c \psi_c+I_2\dot{\delta\chi_c}\psi_c\right],
\end{eqnarray}
where the sound speed of the scalar graviton is given by 
\begin{equation}
    c_s^2=\frac{2c}{3b}\left(1+\frac{2\rho_{\rm rad}}{\rho_{\rm hde}}\right),
\end{equation}
with $\rho_{\rm rad}=\lambda/(2a^4)$ and 
$\rho_{\rm hde}=c/(a^2\varphi^2)$. The latter is plotted in Fig.~\ref{f:cs}, where we can see that it is positive definite during the inflationary phase. The mass of the scalar mode reads
\begin{eqnarray}
   \nonumber M_s^2&=&2H^2\left[\frac{6c}{b}+\frac{2c-b}{b H R_h}-\frac{-6c^2+6b+cb}{6bH^2R_h^2}-\frac{c^2}{9H^4R_h^4}+\Omega_{\rm rad}\left(1+\frac{8c}{b}-\frac{2}{H R_h}-\frac{4c}{3H^2R_h^2}\right) \right .\\
   \nonumber &+&\left.\Omega_{\rm rad}^2\left(\frac{4c}{b\Omega_{\rm hde}}-4\right) \right]+\dot{\chi}^2\left[\frac{1}{6}-\frac{6 H^2 R_h^2 \Omega_{\rm hde}}{b}\left(1+\frac{2\Omega_{\rm rad}}{3}\right)-\frac{\Omega_{\rm hde}}{2}\left(1+\frac{2\Omega_{\rm rad}}{\Omega_{\rm hde}}\right)\right.\\
    &-&\left.\frac{R_h^2 H^2 \Omega_{\rm hde}^2}{b }\left(1+\frac{4\Omega_{\rm rad}^2}{\Omega_{\rm hde}^2}\right)\right]-\frac{2V}{3},
\end{eqnarray}
while for the inflaton term we have
\begin{eqnarray}
   M_{\chi}^2=2H^2\left(-1+\Omega_{\rm rad}+\frac{\Omega_{\rm hde}}{2}\right)+\frac{2V_{,\chi}\dot{\chi}}{H}+\frac{7\dot{\chi}^2}{2 }-\frac{\Omega_{\rm hde}\dot{\chi}^2}{2 }\left(1+\frac{3b}{2c}\right)-\Omega_{\rm rad}\dot{\chi}^2-\frac{\dot{\chi}^4}{2H^2}+V_{,\chi\chi},
\end{eqnarray}
with $R_h\equiv a\varphi$, 
$\Omega_{\rm rad}=\rho_{\rm rad}/(3H^2)$ and 
$\Omega_{\rm hde}=\rho_{\rm hde}/(3H^2)$. For the coefficients of the interaction terms we find
\begin{eqnarray} \nonumber I_1&=&\left\{-4\left[3ba^3H\varphi-(3b+8c)a^4H^2\varphi^2+b\lambda \varphi^2+a^2(bc-2H^2\lambda \varphi^4)\right]\dot{\chi}\right.\\
&-&\left.a^2\varphi^2\left[(3b+2c)a^2+2\lambda \varphi^2\right]\dot{\chi}^3+2a^2H\varphi^2\left[(3b+2c)a^2+2\lambda \varphi^2\right]V_{,\chi}\right\}\left\{ 6 a^4 H^2  \varphi^3 \sqrt{b}\right\}^{-1}
\end{eqnarray}
and
\begin{eqnarray}
   I_2=\frac{\left[(2c-3b)a^2+2\lambda\varphi^2\right]\dot{\chi}}{3a^2H\varphi\sqrt{b}}.
\end{eqnarray}

From the background evolution we know that the inflaton density dominates over the dark sector, namely $\Omega_{\rm rad}$ and $\Omega_{\rm hde}$. Therefore, in order to simplify the quadratic action, we rewrite the Lagrangian multiplier $\lambda$ and the time-like Stueckelberg field $\varphi$ in terms of the fractional densities and expand up to zero order in $\Omega_{\rm rad}$ and $\Omega_{\rm hde}$. We then end up with the following action:
\begin{eqnarray}\label{quadratic scalar}
 \nonumber  S_{\rm scalar}^{(2)}&=&\int dt d^3 k \left\{\frac{1}{2}a\dot{\delta\chi_c}^2+\frac{1}{2}a\dot{\psi_c}^2-\left[\frac{ck^2}{3ba}\left(1+\frac{2\Omega_{\rm rad}}{\Omega_{\rm hde}}\right)+\frac{H^2a(6c-b)}{b}+\frac{a\dot{\chi}^2(b-4c)}{4b }\right]\psi_c^2\right.\\
   &+&\left.\left[-\frac{k^2}{2a}+\frac{a}{4}\left(4H^2-\frac{4V_{,\chi}\dot{\chi}}{H }-7\dot{\chi}^2+\frac{\dot{\chi}^4}{H^2 }-2V_{,\chi\chi}\right)\right]\delta\chi_c^2\right\},
\end{eqnarray}
where the graviton $\psi_c$, which corresponds to the scalar degree of the HDE, 
and the inflaton perturbation $\delta\chi_c$ decouple. With the above expression we compute the scalar power spectrum in section \ref{scalar power spectrum}.

\subsection{Tensor perturbations}
The quadratic action for the tensor perturbation, obtained from \eqref{action}, is given by
\begin{equation}
    S_{\rm tensor}^{(2)}=\frac{1}{8}\int d\tau d^3 k a^2\left[ {\gamma}'^{ij}{\gamma}'_{ij}-(k^2+M_{GW}^2 a^2) \gamma_{ij}\gamma^{ij}\right],
\end{equation}
where $\tau$ is the conformal time, prime denotes derivative with respect to $\tau$ and the mass of the tensor mode reads
\begin{equation}\label{mgw}
    M_{\rm GW}^2=\frac{6c-b+12H^2R_h^2\Omega_{\rm rad}}{6 R_h^2}.
\end{equation}

We then define the  normalized Mukhanov variable as 
$v_k\equiv a \gamma_k/2$and rewrite the action in the form
\begin{equation}\label{lagrangian tensor}
    S_{\rm tensor}^{(2)}=\int d\tau d^3 k\frac{1}{2}\left[ {v_k}'^2+v_k^2\left(-k^2-\frac{\lambda}{3 a^2}+\frac{b-6c}{6 \varphi^2}+\frac{{a}''}{a}\right)\right],
\end{equation}
which allows us to compute the power spectrum in section \ref{tensor power spectrum}.

\section{Power spectra}\label{Power spectra}
\subsection{Scalar power spectrum}\label{scalar power spectrum}
As we have seen in section \ref{scalar perturbations}, the graviton $\psi_c$ and the inflaton perturbation $\delta\chi_c$ decouple when one considers an expansion of the action for low values of $\Omega_{\rm rad}$ and $\Omega_{\rm hde}$, which is corroborated by the results for background evolution. In this case, we are able to compute the power spectrum for $\psi$ and $\delta\chi$ separately.  

\subsubsection{Power spectrum of the inflaton perturbation}\label{power spectrum inflaton}
Let us start by computing the power spectrum related to $\delta\chi_c$. From the decoupled quadratic action \eqref{quadratic scalar} we have the action corresponding to the inflaton perturbation, which reads
\begin{equation}
    S_{\delta\chi_c}^{(2)}=\int d\tau d^3 k \left[\frac{{\delta\chi_c}'^2}{2}-\frac{k^2}{2}\delta\chi_c^2+F(\tau)\delta\chi_c^2\right]
\end{equation}
in terms of the conformal time $\tau$, where $F(\tau)$ 
is given by
\begin{eqnarray}
   F(\tau)=\frac{1}{4}\left[-\frac{4aV_{,\chi}{\chi}'}{H }-7{\chi}'^2+\frac{{\chi}'^4}{a^2H^2}+a^2\left(4H^2-2V_{,\chi\chi}\right)\right].
\end{eqnarray}

The corresponding equation of motion reads
\begin{equation}
    {\delta\chi_{ck}}''+\left[k^2-2F(\tau)\right]\delta\chi_{ck}=0,
\end{equation}
which is simplified to
\begin{align}\label{eqdifchisr}
    \delta \chi_c^{\prime\prime} + \left[k^2 -2 a^2H^2 \left(1+ \epsilon - \frac{3}{2} \eta\right) \right] \delta \chi_c=0
\end{align}
up to first order in slow-roll approximation, where $\epsilon$ and $\eta$ are the slow-roll parameters given by 
\begin{equation}
    \epsilon\equiv -\frac{\dot{H}}{H^2}\simeq \frac{\dot{\chi}^2}{2H^2},\qquad \eta\equiv -\frac{\ddot{\chi}}{H\dot{\chi}},
\end{equation}
and we have used $V\simeq 3 H^2$. 

Defining $x\equiv k/(aH)$ we obtain
\begin{equation}
    x^2(1-2\epsilon)\frac{d^2\delta\chi_c}{dx^2}+\left[x^2-2\left(1+\epsilon-\frac{3}{2}\eta\right)\right]\delta\chi_c=0,
\end{equation}
which leads 
to the following leading order solution:
\begin{equation}\label{deltachi2}
    \delta\chi_c=-\frac{1}{2}\sqrt{\frac{\pi x}{k}}\left[J_\nu(x)+iY_\nu(x)\right],
\end{equation}
where $J_\nu$ and $Y_\nu$ are Bessel functions of the first and second kind, respectively, and
\begin{equation}
    \nu\simeq \frac{3}{2}+2\epsilon-\eta
\end{equation}
up to first order in slow-roll approximation.
The initial conditions were fixed in order to obtain the commutation relation between the annihilation and creation operators and the Bunch-Davis vacuum in the infinite past.

Considering the solution \eqref{deltachi2}, we obtain the corresponding solution to $\delta\chi_k$ via equation \eqref{psic}, which, in terms of 
$x$ and the Hankel function of the first kind $H^{(1)}_\nu$, reads
\begin{equation}
    \delta\chi_k=-\frac{1}{2a}\sqrt{\frac{\pi x}{k}}H^{(1)}_\nu(x).
\end{equation}

In the superhorizon regime, i.e. $x\ll 1$, the two-point correlation function of $\delta\chi$ is given by
\begin{equation}
    \left \langle  \delta\chi_k(\tau)\delta\chi_{{k}'}(\tau) \right \rangle=(2\pi)^3\delta(\vec{k}+\vec{{k}'})\frac{4^{-1+\nu}x^{1-2\nu}\Gamma(\nu)^2}{k \pi a^2}
\end{equation}
where $\Gamma$ represents the gamma function. Therefore the dimensionless power spectrum reads 
\begin{equation}\label{psdchi}
    \Delta_{\delta\chi}^2=\frac{2^{-3+2\nu}x^{3-2\nu}\Gamma(\nu)^2 H^2}{\pi^3}\biggl| \biggr._{k=aH},
\end{equation}
where $k=aH$ indicates that it must be evaluated at horizon crossing. Therefore, the scalar spectral index is obtained as
\begin{equation}
    n_s-1=3-2\nu=4\epsilon-2\eta,
\end{equation}
which corresponds to the standard single field result.

\begin{figure}
\begin{center}
\includegraphics[width=7cm]{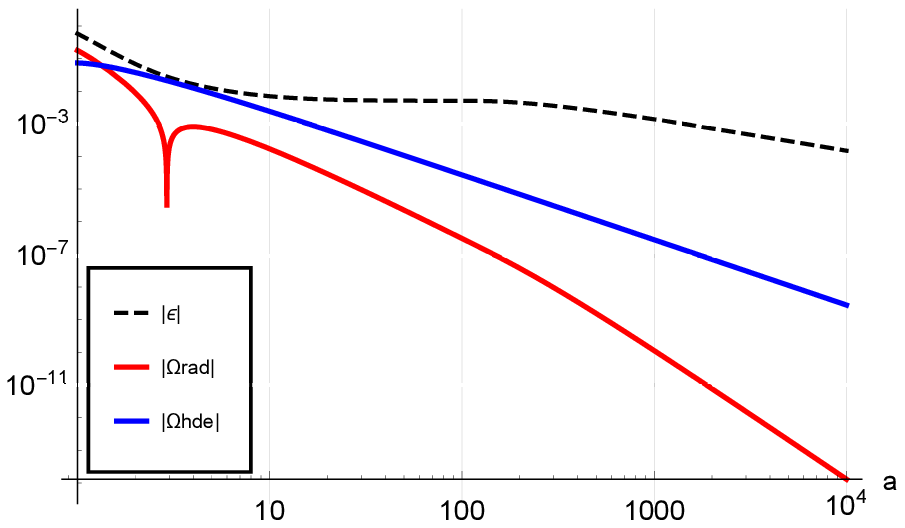}
\includegraphics[width=7cm]{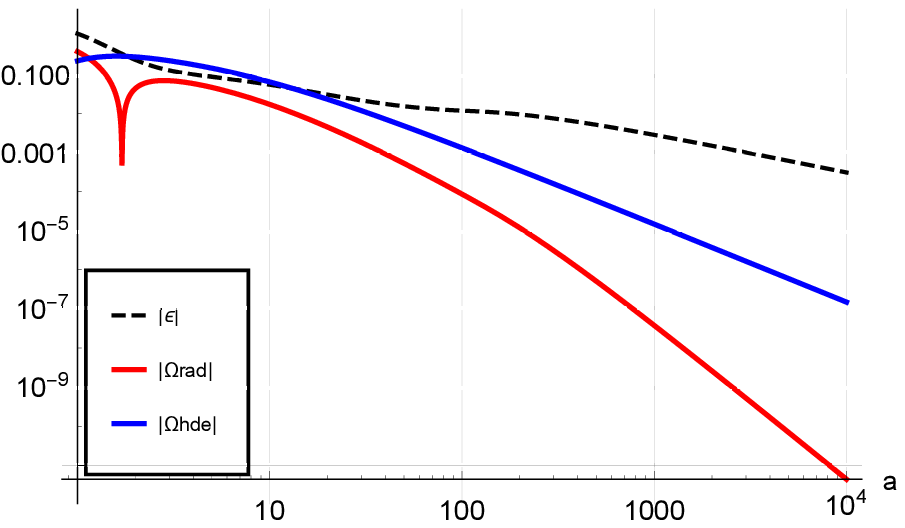}
\end{center}
\caption{\label{epsilonandomegas} Comparison between the dark densities $\Omega_{\rm rad}$ and $\Omega_{\rm hde}$ and the slow-roll parameter $\epsilon$. Here $c=2.2$, $\tilde{m}=0.1$, $\tilde{\varphi}_i=\frac{1}{\sqrt{0.1}}$, $\tilde{\rho}_i=0.75$ and $\tilde{V}=0.7$ (left) and $c=2.2$, $\tilde{m}=0.1$, $\tilde{\varphi}_i=\frac{1}{\sqrt{0.3}}$, $\tilde{\rho}_i=0.4$ and $\tilde{V}=0.36$ (right).}
\end{figure}

The next-to-leading order contribution to the 
equation of motion is of order $\sqrt{\Omega_i}$, $i={\rm hde},{\rm rad}$, and can be considered as a source term. 
In the slow-roll regime we find
\begin{align}
    \delta \chi_c^{\prime\prime} + (k^2- 2 a^2H^2) \delta \chi_c = \psi_c \frac{a^2H^2 \sqrt{\epsilon}}{\sqrt{6 c b \Omega_{hde}}} \left[ (10c-3b) \Omega_{hde} + 4c \Omega_{rad} \right] + \psi_c^\prime \frac{aH \sqrt{\epsilon}}{\sqrt{6 c b \Omega_{hde}}}\left[ (3b-2c) \Omega_{hde} - 4c \Omega_{rad} \right].
\end{align}
Therefore in the equation of motion this contribution is not only suppressed by the square root of the fractional densities related to the holographic dark energy, but also by $\sqrt{\epsilon}$. For the power spectrum this means a correction of order $\Omega_i \epsilon$, which is even smaller than the slow-roll correction of order $\epsilon^2$, as one can see in Fig.~\ref{epsilonandomegas}.

\subsubsection{Power spectrum of the graviton}
Now let us compute the power spectrum related to the graviton $\psi$. From the quadratic action \eqref{quadratic scalar} we have the following action for $\psi_c$ in terms of the conformal time $\tau$:
\begin{equation}
    S_{\psi_c}^{(2)}=\int d\tau d^3k \left[ \frac{{\psi_c}'^2}{2}-\frac{k^2c_s^2}{2}\psi_c^2+G(\tau)\psi_c^2 \right],
\end{equation}
where prime denotes derivative with respect to $\tau$ and 
\begin{equation}
    G(\tau)=\frac{4(b-6c)a^2 H^2-(b-4c){\chi}'^2}{4b}
\end{equation}

The corresponding equation of motion is given by
\begin{equation}
    {\psi_{ck}}''+\left[k^2c_s^2-2G(\tau)\right]\psi_{ck}=0,
\end{equation}
which, when considering slow-roll approximation up to first order, reduces to
\begin{equation}
    {\psi_{ck}}''+\left[k^2 c_s^2-2a^2H^2+\frac{12ca^2H^2+\epsilon(b-4c)a^2H^2}{b}\right] \psi_{ck}=0.
\end{equation}
Defining $x\equiv c_s k/(aH)$ and $s\equiv \dot c_s/(H c_s)$ we rewrite the equation of motion as
\begin{equation}
    x^2(1-2\epsilon)\frac{d^2 \psi_{ck}}{dx^2}-s x \frac{d\psi_{ck}}{dx}+\left[-2+x^2-\frac{4c(-3+\epsilon)}{b}+\epsilon\right]\psi_{ck}=0,
\end{equation}
which leads to the following solution:
\begin{equation}
    \psi_{ck}=-\frac{1}{2}\sqrt{\frac{\pi}{2}}\left[\frac{x(2-4\epsilon)}{c_s k}\right]^{\frac{1+s-2\epsilon}{2-4\epsilon}}\left[J_\nu(x)+iY_\nu(x)\right],
\end{equation}
where
\begin{equation}
    \nu=-\frac{\sqrt{-16c(3-7\epsilon)+b(9+2s-24\epsilon)}}{2\sqrt{b}(-1+2\epsilon)}.
\end{equation}
If $\nu$ is imaginary, the power spectrum is highly suppressed, since in this case the solution is in terms of hyperbolic trigonometric functions that decay instead of oscillating. 
Therefore, from now on we assume that $\nu$ is real.

Recalling the definition of the canonical variable given by \eqref{psic}, we have the expression for the graviton $\psi$, with which we calculate the power spectrum. In terms of the Hankel function of the first kind $H^{(1)}_{\nu}$ and the variable $x\equiv-c_s k\tau$, the solution reads
\begin{equation}
    \psi_k=-\frac{H\varphi}{2}\sqrt{\frac{\pi x}{c_s k b}}H^{(1)}_{\nu}(x),
\end{equation}
where we have neglected the parameters $\epsilon$ and $s$ in the exponent.

The two-point correlation function of the variable $\psi$ in the superhorizon regime, i.e. $x\ll 1$, is given by
\begin{equation}
    \left \langle\psi_k(\tau)\psi_{{k}'}(\tau)  \right \rangle=(2\pi)^3\delta(\vec{k}+\vec{{k}'})\frac{4^{-1+\nu}x^{1-2\nu}\Gamma(\nu)^2H^2\varphi^2}{b c_s k\pi},
\end{equation}
which corresponds to the power spectrum
\begin{equation}\label{pspsi}
    \Delta_{\psi}^2=\frac{2^{-3+2\nu}x^{3-2\nu}\Gamma(\nu)^2a^2 H^4\varphi^2}{b\pi^3 c_s^3}\biggl| \biggr._{c_S k=aH},
\end{equation}
to be evaluated at horizon crossing.

\subsubsection{Comoving curvature power spectrum}\label{Comoving curvature power spectrum}

In order to obtain the comoving curvature
perturbation, which is given by 
\begin{eqnarray}
    R= -\psi-\frac{H}{\rho+p}\delta q,
\end{eqnarray}
we compute $\delta q$ via $\partial_i\delta q\equiv\delta T_i^0$, where $T_{\mu\nu}$ is the energy-momentum tensor.
Perturbing the latter, which encompasses the inflaton and the holographic and radiation terms, we find\footnote{Note that in this step $\delta\chi$ corresponds to the original perturbation, i.e. the one defined previously to the re-scaling \eqref{deltachioriginal}.}
\begin{eqnarray}
   \delta q&=&-\dot{\chi}\delta\chi,
\end{eqnarray}
while the total density and pressure read respectively
\begin{eqnarray}
    \rho&=&\frac{1}{2}\dot{\chi}^2+V+\frac{c}{a^2\varphi^2}+\frac{\lambda}{2a^4},\\
    p&=&\frac{1}{2}\dot{\chi}^2-V-\frac{c}{3a^2\varphi^2}+\frac{\lambda}{6a^4}.
\end{eqnarray}
Therefore, the curvature perturbation is given by
\begin{equation}\label{Rexpression}
    R=\frac{\dot{\chi}}{2 \epsilon a H}\delta\chi_c -\sqrt{\frac{c}{3b\Omega_{\rm hde}}}\frac{\Omega_{\rm hde}+2\Omega_{\rm rad}}{\epsilon a}\psi_c,
\end{equation}
where we have used both the re-scaling \eqref{deltachioriginal} and the canonical transformation \eqref{psic}.

As a result, the power spectrum reads 
\begin{eqnarray}
  \nonumber  \Delta_R^2&=&\frac{\dot{\chi}^2}{4\epsilon^2 a^2H^2}\Delta^2_{\delta\chi_c}+\frac{c(\Omega_{\rm hde}+2\Omega_{\rm rad})^2}{3b\epsilon^2 a^2\Omega_{\rm hde}}\Delta^2_{\psi_c}\\
  &=&\frac{1}{2\epsilon a^2}\left(1-\frac{\Omega_{\rm hde}+2 \Omega_{\rm rad}}{\epsilon}\right)\Delta^2_{\delta\chi_c}+\frac{c(\Omega_{\rm hde}+2\Omega_{\rm rad})^2}{3\epsilon^2 ba^2\Omega_{\rm hde}}\Delta^2_{\psi_c}.
\end{eqnarray}
Note that the contribution coming from $\psi_c$ decays away, as the dark radiation and the holographic components in the prefactor are dominated by the inflaton. Therefore, although $\Delta^2_{\psi_c}$ is approximately frozen after horizon crossing, the overall contribution becomes extremely small.
Note also that \eqref{Rexpression} corresponds to the comoving curvature power spectrum at the end of inflation, which is not conserved in the presence of entropy perturbations. However, since the second term in the right hand size of \eqref{Rexpression} decays rapidly, becoming negligible after the very first e-folds, the majority of modes is approximately frozen after horizon crossing. The modes that leave the horizon during the first e-folds might still evolve, but they correspond to the lowest multipoles in the CMB power spectrum, which are not tightly constraint due to cosmic variance. Finally, in terms of \eqref{psdchi} and \eqref{pspsi}, we have
\begin{equation}\label{power spectrum R2}
    \Delta_R^2=\frac{1}{2\epsilon }\left(1-\frac{\Omega_{\rm hde}+2 \Omega_{\rm rad}}{\epsilon}\right)\Delta^2_{\delta\chi}+\frac{c(\Omega_{\rm hde}+2\Omega_{\rm rad})^2}{3\epsilon^2 a^2 H^2 \varphi^2 \Omega_{\rm hde}}\Delta^2_{\psi}\simeq \frac{1}{2\epsilon}\Delta^2_{\delta\chi}.
\end{equation}

\subsection{Tensor power spectrum}\label{tensor power spectrum}
The quadratic action \eqref{lagrangian tensor} leads to the following equation of motion:
\begin{equation}\label{eq of motion}
   {v_k}''+v_k\left( k^2+\frac{\lambda}{3 a^2}-\frac{b}{6\varphi^2}+\frac{c }{\varphi^2}-\frac{ {a}''}{a}\right)=0.
\end{equation}

In a pure de Sitter universe, 
it is rewritten as 
\begin{equation}\label{Mukh eq}
    {v_k}''+\left( k^2-\frac{2}{\tau^2}+M^2\right) v_k=0,
\end{equation}
where $M$ is approximately constant\footnote{Note from equation \eqref{mgw} that the dominant terms in the mass of the tensor mode are proportional to $a^{-2}$, while $\Omega_{\rm rad}$ decays faster, as one can see in Fig.~\ref{epsilonandomegas}. Therefore, after a few e-folds, $M^2=M_{GW}^2a^2$ is approximately constant.} and related to the mass of the tensor mode \eqref{mgw} as
\begin{equation}
    M^2=M_{\rm GW}^2 a^2=\frac{\lambda H^2\tau^2}{3 }-\frac{b}{6\varphi^2}+\frac{c}{\varphi^2}.
\end{equation}

The solution satisfying both equation \eqref{Mukh eq} and the commutation relation for the annihilation and creation operators, i.e. $[\hat{a}_{\vec{k}},\hat{a}_{\vec{k}'}^\dagger]=(2\pi)^3\delta(\vec{k}-\vec{k}')$, reads
\begin{equation}
    v_k=\frac{\operatorname{e}^{- i \sqrt{k^2+M^2}\tau}}{\sqrt{2}(k^2+M^2)^{\frac{1}{4}}}\left( 1-\frac{i}{\sqrt{k^2+M^2}\tau} \right).
\end{equation}

\begin{figure}
\begin{center}
\includegraphics[width=7cm]{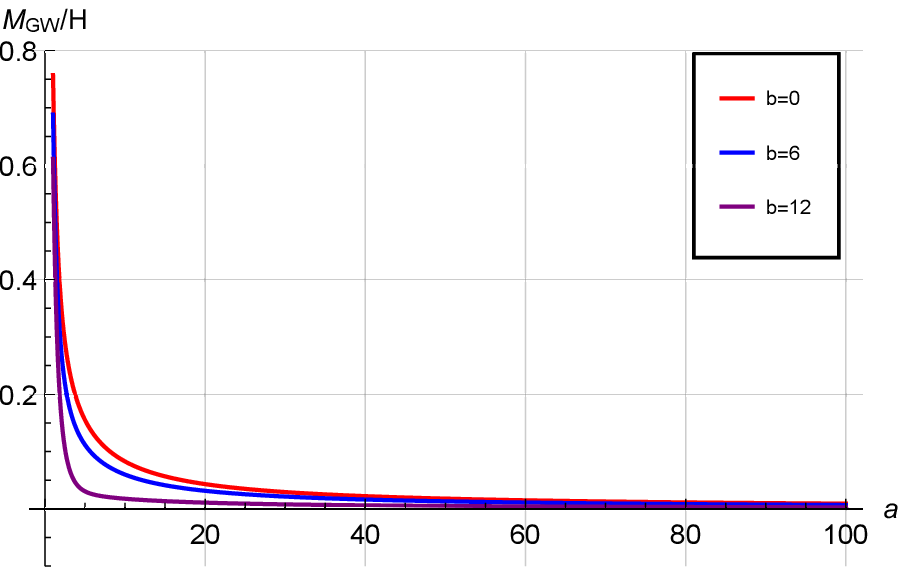}
\includegraphics[width=7cm]{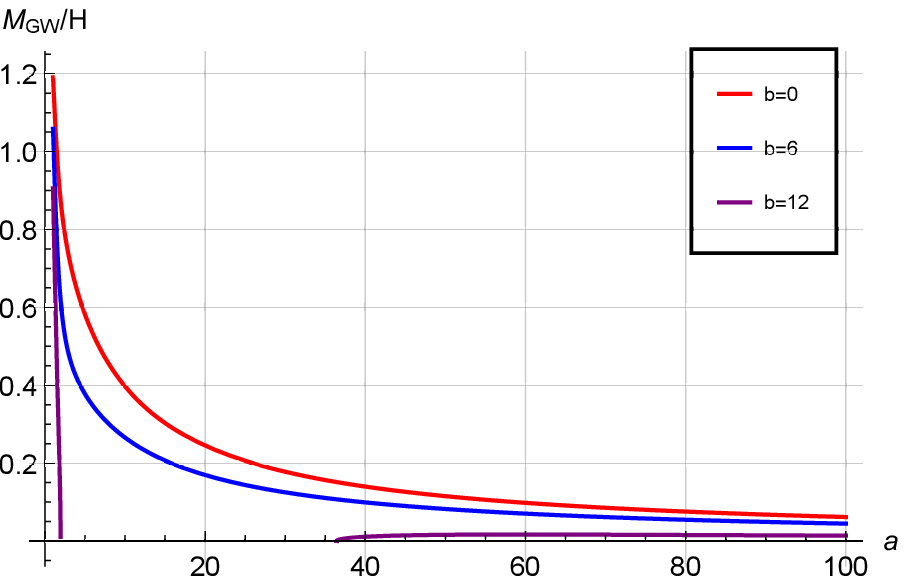}
\end{center}
\caption{\label{f:MGW} $M_{GW}/H$ for $c=2.2$. Here $\tilde{m}=0.1$, $\tilde{\varphi}_i=\frac{1}{\sqrt{0.1}}$, $\tilde{\rho}_i=0.75$ and $\tilde{V}=0.7$ (left) and $c=2.2$, $\tilde{m}=0.1$, $\tilde{\varphi}_i=\frac{1}{\sqrt{0.3}}$, $\tilde{\rho}_i=0.4$ and $\tilde{V}=0.36$ (right).}
\end{figure}

The two-point correlation function of the tensor perturbation $\gamma_{k}$ is readily obtained as
\begin{eqnarray}
   \nonumber \left \langle \gamma_k \gamma_k' \right \rangle &=& (2\pi)^3\delta(k+k')4\frac{|v_k|^2}{a^2}\\
    &=&16\pi^3\delta(k+k')\frac{H^2(1+k^2\tau^2+M^2\tau^2)}{(k^2+M^2)^{\frac{3}{2}}},
\end{eqnarray}
leading to the following dimensionless power spectrum:
\begin{equation}
   \Delta_t^2= 2\Delta_\gamma^2=2\frac{H^2 k^3 (1+k^2\tau^2+M^2\tau^2)}{\pi^2 (k^2+M^2)^{\frac{3}{2}}}.
\end{equation}
For superhorizon modes, we have
\begin{equation}\label{super h power spectrum}
   \Delta_t^2= 2\Delta_\gamma^2=2\frac{H^2 k^3 }{\pi^2 (k^2+M^2)^{\frac{3}{2}}} \biggl| \biggr._{\sqrt{k^2+M^2}=aH}=\frac{2 H^2}{\pi^2} \left(1- \frac{M_{GW}^2}{H^2} \right)^{\frac{3}{2}},
\end{equation}
where the second equality must be evaluated at horizon crossing, i.e. $\sqrt{k^2+M^2}=aH$. 
The ratio $M_{GW}/H$ is plotted in figure \ref{f:MGW}, where we see that it decays rapidly.

The spectral index $n_t$ can be obtained from the superhorizon power spectrum \eqref{super h power spectrum} via
\begin{equation}
    n_t=\frac{d\ln{\Delta_t^2}}{d \ln{k}},
\end{equation}
which can be computed
using the chain rule
\begin{equation}
   n_t= \frac{d\ln{\Delta_t^2}}{d N}\frac{dN}{d\ln{k}},
\end{equation}
where $N$ is the number of e-folds. The first factor is obtained directly from the power spectrum, while the second comes from  $\ln{k}=N+\ln{H}+\frac{1}{2} \ln{\left(1-\frac{M^2}{a^2 H^2}\right)}$, leading to
\begin{eqnarray}
  \nonumber  \frac{dN}{d\ln{k}}&=&\left[ 1+\frac{d\ln{H}}{dN}+\frac{1}{2}\frac{d\ln{\left(1-\frac{M^2}{a^2 H^2}\right)}}{dN}\right]^{-1}=\\
    &\simeq& \left[(1-\epsilon)\left(1+\frac{M_{GW}^2}{H^2}\right)-\frac{\epsilon_M M_{GW}^2}{H^2} \right]^{-1},
\end{eqnarray}
up to leading order in $M_{GW}^2/H^2$, where $\epsilon_M\equiv M_{,N}/M$.

Using that $H_{,N}/H=-\epsilon$, where $_{,N}$ represents derivative with respect to $N$, one finds
\begin{eqnarray}
\nonumber n_t&=&-\frac{2\epsilon}{(1-\epsilon)\left(1+\frac{M^2}{a^2 H^2}\right)-\frac{\epsilon_M M^2}{a^2 H^2}}+\frac{3M^2(-\epsilon_M k \tau^2 \sqrt{1-M^2\tau^2}+M\tau^3 M_{,N}+\tau_{,N})}{k\tau^2\sqrt{1-M^2\tau^2}(k^2+M^2)\left[(1-\epsilon)\left(1+\frac{M^2}{a^2H^2}\right)-\frac{\epsilon_M M^2}{a^2 H^2}\right]}=\\
&\simeq & %\frac{-2\epsilon+3 \frac{M_{GW}^2}{H^2}(1-\epsilon_M)}{1+\frac{M_{GW}^2}{H^2}}
-2\epsilon+\frac{M_{GW}^2}{H^2}(3-3\epsilon_M+2\epsilon)
\end{eqnarray}
up to first order in $\epsilon$ and $\epsilon_M$, and up to second order in $M_{GW}/H$. In the last equality we have used that $\tau_{,N}=-(1-\epsilon)\tau$,  $\tau=-1/(aH)$ and $\sqrt{k^2+M^2}=aH$. Therefore, since $M_{GW}\ll H$ from the very beginning of inflation on, we obtain a tensor spectral index similar to the standard one for single field inflation.

\section{Conclusions}\label{Conclusions}
In this paper we analyzed the effects of the two components of the HDE model, namely the holographic and the dark radiation terms, in a single field slow-roll inflation, both in the scalar and tensor sectors.
We started by numerically computing the evolution of the components at the background level, which led to the conclusion that the inflaton dominates after a few e-folds, while the holographic and (especially) the dark radiation terms decay rapidly. Therefore, at the background level, the inflationary phase is not spoiled by the presence of HDE. 

For the scalar perturbation we showed that the two scalar degrees of freedom, i.e. the inflaton perturbation and the graviton, decouple when expanding the quadratic action up to zero order in $\Omega_{\rm hde}$ and $\Omega_{\rm rad}$. The next-to-leading order correction to this expansion is of order $\sqrt{\Omega_{i}\epsilon}$, $i={\rm hde},{\rm rad}$, i.e. a correction suppressed not only by the square root of the tiny fractional densities of the dark sector, but also by the square root of the first slow-roll parameter. The correction to the curvature power spectrum coming from the graviton decays rapidly, as the dark radiation and holographic components are dominated by the inflaton. Moreover, the correction to the inflaton power spectrum also decays as $\Omega_{hde}$ and $\Omega_{rad}$ become smaller than $\epsilon$. 

For the tensor power spectrum we find a dependence on the mass of the tensor mode, which decays rapidly in the beginning of inflation. The resulting power spectrum for $M_{GW}\ll H$ is the usual single field result. 

Taking all the above features into account, we conclude that HDE is compatible with single field slow-roll inflation, being in agreement with the current constraints on the scalar power spectrum, since the extra contributions decay quite fast, and on primordial gravitational waves, since it leads to the usual tensor amplitude. In case of detection of primordial gravitational waves, the tensor spectral index can be used as evidence in favour or against the HDE considered in this work, constraining the mass of the tensor mode.

\begin{acknowledgments}
C.L. and P.C.M.D. are supported by the grant No. UMO-2018/30/Q/ST9/00795 from the National Science Centre, Poland. A.G. receives support by the grant No. UMO-2021/40/C/ST9/00015 from the National Science Centre, Poland. 
\end{acknowledgments}

%\appendix


\begin{thebibliography}{99}
\bibitem{Riess} 
A. G. Riess {\it et al}, AJ {\bf 116}, 1009 (1998).

\bibitem{Perlmutter} 
S. Perlmutter {\it et al}, ApJ {\bf 517} 565 (1999).

\bibitem{PeeblesDE} 
P. J. E. Peebles and B. Ratra, Rev. Mod. Phys. {\bf 75} 559 (2003).

\bibitem{WMAP} 
G. Hinshaw {\it et al}, ApJS {\bf 208} 19 (2013).

\bibitem{Planck2013}
Planck Collaboration, A$\&$A {\bf 571} A16 (2014).

\bibitem{Planck2015}
Planck Collaboration, A$\&$A {\bf 594} A13 (2016).

\bibitem{Planck2018}
Planck Collaboration, A$\&$A {\bf 641} A6 (2020).

\bibitem{LSS}
David Parkinson {\it et al}, Phys. Rev. D {\bf 86} 103518 (2012).

\bibitem{SWE1}
S. Ho, C. Hirata, N. Padmanabhan, U. Seljak, and N. Bahcall, Phys. Rev. D {\bf 78} 043519 (2008).

\bibitem{SWE2}
T. Giannantonio, R. Scranton, R. G. Crittenden, R. C. Nichol, S. P. Boughn, A. D. Myers, and G. T. Richards, Phys. Rev. D {\bf 77} 123520 (2008).

\bibitem{measurementsHubble}
C. Ma and T. Zhang, ApJ {\bf 730} 74 (2011).

\bibitem{DES1}
T. M. C. Abbott {\it et al} (DES Collaboration), Phys. Rev. D {\bf 102} 023509 (2020).

\bibitem{DES3}
T. M. C. Abbott {\it et al} (DES Collaboration), arXiv:2105.13549 [astro-ph.CO] (2021).

\bibitem{HP1} 
G. ´t Hooft, arXiv:gr-qc/9310026 (1993).

\bibitem{HP} 
L. Susskind, Journal of Mathematical Physics {\bf 36} 6377 (1995).

%\cite{Cohen:1998zx}
\bibitem{Cohen:1998zx}
A.~G.~Cohen, D.~B.~Kaplan and A.~E.~Nelson,
%``Effective field theory, black holes, and the cosmological constant,''
Phys. Rev. Lett. \textbf{82} (1999), 4971-4974
doi:10.1103/PhysRevLett.82.4971
[arXiv:hep-th/9803132 [hep-th]].
%1058 citations counted in INSPIRE as of 12 Oct 2021

%\cite{Weinberg:1988cp}
\bibitem{Weinberg:1988cp}
S.~Weinberg,
%``The Cosmological Constant Problem,''
Rev. Mod. Phys. \textbf{61} (1989), 1-23
doi:10.1103/RevModPhys.61.1
%5086 citations counted in INSPIRE as of 12 Oct 2021


\bibitem{MiaoLi1} 
M. Li, Phys. Lett. B {\bf 603} 1-5 (2004).


\bibitem{MiaoLi2} 
S. Wang, Y. Wang and M. Li, Physics Reports {\bf 696} 1-57 (2017).

\bibitem{MiaoLi3} 
M. Li, R. Miao, arXiv:1210.0966[hep-th] (2012).

\bibitem{ChunshanHDE} 
C.~Lin,
%``An effective field theory of holographic dark energy,''
JCAP07(2021)003  
%doi:10.1088/1475-7516/2021/07/003
[arXiv:2101.08092 [hep-th]].
%2 citations counted in INSPIRE as of 15 Oct 2021
%\cite{Ganz:2021hmp}
\bibitem{Ganz:2021hmp}
A.~Ganz and C.~Lin,
%``Structure Formation in the Effective Field Theory of Holographic Dark Energy,''
[arXiv:2109.07420 [gr-qc]].
%0 citations counted in INSPIRE as of 15 Oct 2021

\bibitem{Inflation1} 
A. H. Guth, Phys. Rev. D {\bf 23} 347 (1981).

\bibitem{Inflation2} 
A. Albrecht and P. J. Steinhardt, Phys. Rev. Lett. {\bf 48} 1220 (1982).

\bibitem{Inflation3} 
A. D. Linde, Phys. Lett. B {\bf 108} 389-393 (1982).

\bibitem{Inflation4} 
A. D. Linde, Phys. Lett. B {\bf 129} 177-181 (1983).

\bibitem{Inflation5} 
A. A. {Starobinski{\v{i}}}, Soviet Journal of Experimental and Theoretical Physics Letters {\bf 30} 682 (1979).

\bibitem{InflationHDE} 
B. Chen, M. Li and Y. Wang, Nuclear Physics B {\bf 774} 1-3 (2007).

\bibitem{HDEinflationdifferent} 
S. Nojiri, S. D. Odintsov and E. N. Saridakis, Phys. Lett. B {\bf 797} 134829 (2019).

\bibitem{MiaoLiHDE} 
M. Li, X. Li, J. Meng and Z. Zhang, Phys. Rev. D {\bf 88} 023503 (2013).

\end{thebibliography}
\end{document}